\documentclass[aps,prb,showpacs,twocolumn]{revtex4-1}
\usepackage{amssymb}
\usepackage{amsmath}
\usepackage{graphicx}
\usepackage{dsfont}
\usepackage{times}

\begin{document}

\title{Higher-order level spacings in random matrix theory based on Wigner's conjecture}
\author{Wen-Jia Rao$^1$}
\email{wjrao@hdu.edu.cn}
\affiliation{School of Science, Hangzhou Dianzi University, Hangzhou 310027, China.}

\begin{abstract}
The distribution of higher order level spacings, i.e. the distribution of $%
\{s_{i}^{(n)}=E_{i+n}-E_{i}\}$ with $n\geq 1$ is derived analytically using
a Wigner-like surmise for Gaussian ensembles of random matrix as well
as Poisson ensemble. It is found $s^{(n)}$ in Gaussian ensembles follows a
generalized Wigner-Dyson distribution with rescaled parameter $\alpha=\nu
C_{n+1}^2+n-1$, while that in Poisson ensemble follows a generalized
semi-Poisson distribution with index $n$. Numerical evidences are provided
through simulations of random spin systems as well as non-trivial zeros of
Riemann zeta function. The higher order generalizations of gap ratios are
also discussed.
\end{abstract}

\maketitle

\section{Introduction}

\label{intro}

Random matrix theory (RMT) was introduced half a century ago when dealing
with complex nuclei\cite{Porter}, and since then has found various
applications in fields ranging from quantum chaos to isolated many-body
systems\cite{RMP,PR}. This roots in the fact that RMT describes universal
properties of random matrix that depend only on its symmetry while
independent of microscopic details. Specifically, the system with time
reversal invariance is represented by matrix that belongs to the Gaussian
orthogonal ensemble (GOE); the system with spin rotational invariance while
breaks time reversal symmetry belongs to the Gaussian unitary ensemble
(GUE); while Gaussian symplectic ensemble (GSE) represents systems with time
reversal symmetry but breaks spin rotational symmetry.

Among various statistical quantities, the most widely used one is the
distribution of nearest level spacings $\left\{ s_{i}=E_{i+1}-E_{i}\right\} $%
, i.e. the gaps between adjacent energy levels, which measures the strength
of level repulsion. The exact expression for the $P\left( s\right) $ can be
derived analytically for random matrix with large dimension, which is
cumbersome\cite{Mehta,Haake2001}. Instead, for most practical purposes it's
sufficient to employ the so-called Wigner surmise\cite{Wigner} that deals
with $2\times 2$ matrix (this will be reviewed in Sec.~\ref{nearest}), the
out-coming result for $P(s)$ has a neat expression that contains a
polynomial part accounting for level repulsion and an Gaussian decaying part
(see Eq.~(\ref{equ:nearest})).

Different models may and usually do have different density of states (DOS),
hence to compare the universal behavior of level spacings, an unfolding
procedure is required to erase the model dependent information of DOS. To
overcome this obstacle, Oganesyan and Huse\cite{Oganesyan} proposed a new
quantity to study the level statistics, i.e. the ratio between adjacent gaps
$\left\{ r_{i}=s_{i+1}/s_{i}\right\} $, whose distribution $P\left( r\right)
$ is later analytically derived by Atas \textit{et al.}\cite{Atas}. The gap
ratio is independent of local DOS and requires no unfolding procedure
(provided the DOS does not vary in the scale of the spacings involve), hence
has found various applications, especially in the context of many-body
localization (MBL)\cite%
{Huse1,Huse2,Huse3,Sarma,Lev,Agarwal,Luitz,Avishai2002,Regnault16,Regnault162}%
.

Both the nearest level spacing and gap ratio account for the short range
level correlations. However, long range correlations are also important,
especially when studying the MBL transition phenomena. Indeed, there're
several effective models describing the level distribution at the MBL
transition region. For example, the Rosenzweig-Porter model\cite{Shukla},
mean field plasma model\cite{Serbyn}, short-range plasma models (SRPM)\cite%
{SRPM} and its generalization -- so-called weighed SRPM\cite{Sierant19},
Gaussian $\beta $ ensemble\cite{Buijsman} and the generalized $\beta -h$
model\cite{Sierant20}. All of these models more or less describe the
short-range level correlations in the MBL transition region well, and their
difference can only be revealed when long-range correlations are concerned.
For a comparison of these models in describing MBL transition point, see
Ref.~[\onlinecite{Sierant19}].

Commonly, the long-range correlations in a random matrix can be described by
the number variance $\Sigma ^{2}$ or the Dyson-Mehta $\Delta _{3}$ statistics%
\cite{Haake2001}, however, both of them are very sensitive to the concrete
unfolding strategy and have already been a source of misleading signatures%
\cite{Gomez2002}. Instead, it's more direct and numerically easier to study
the higher order level spacings and gap ratios. There're existing works that
generalize the level spacing and gap ratios to higher order, as well as
their applications in studying MBL transitions\cite%
{Sierant19,Sierant20,Tekur1,Tekur,Atas2,Chavda,Magd,Duras,Rubah}. However,
most of these works are numerical or phenomenological, and an analytical
derivation for the distribution of level spacing/gap ratio is still lacking.
Given the importance of higher-order level correlations, it's desirable to
have an analytical formula for them, it is then the purpose of this work to
fill in this gap.

In this work, by using a Wigner-like surmise, we succeeded in obtaining an
analytical expression for the distribution of higher order spacing $\left\{
s_{i}^{\left( n\right) }=E_{i+n}-E_{i}\right\} $ in all the Gaussian
ensembles of RMT, as well as the Poisson ensemble. The results show the
distribution of $s_{i}^{\left( n\right) }$ in the former class follows a
generalized Wigner-Dyson distribution with rescaled parameter; while in
Poisson ensemble it follows a generalized semi-Poisson distribution with
index $n$. Interestingly, the rescaling behavior of higher-order level
spacing is identical to that of the high-order gap ratio found numerically
in Ref.~[\onlinecite{Tekur}], for which we will provide a heuristic
explanation.

This paper is organized as follows. In Sec.~\ref{nearest} we review the
Wigner surmise for deriving the distribution of nearest level spacings, and
present numerical data to validate this surmise. In Sec.~\ref{analytical} we
present the analytical derivation for higher order level spacings using a
Wigner-like surmise, and numerical fittings are given in Sec.~\ref{numerics}%
. In Sec.~\ref{ratio} we discuss the generalization of gap ratios to higher
order. Conclusion and discussion come in Sec.~\ref{conclusion}.

\section{Nearest Level Spacings}

\label{nearest}

We begin with the discussion about nearest level spacings, our starting
point probability distribution of energy levels $P\left( \left\{
E_{i}\right\} \right) $ in three Gaussian ensembles, whose expression can be
found in any textbook on RMT (e.g. Ref.~[\onlinecite{Haake2001}]),%
\begin{equation}
P\left( \left\{ E_{i}\right\} \right) \propto \prod_{i<j}\left\vert
E_{i}-E_{j}\right\vert ^{\nu }e^{-A\sum_{i}E_{i}^{2}}  \label{equ:Dist}
\end{equation}%
where $\nu =1,2,4$ for GOE,GUE,GSE respectively. The distribution of nearest
level spacing can then be written as
\begin{equation}
P\left( s\right) =\int \prod_{i=1}^{N}dE_{i}P\left( \left\{ E_{i}\right\}
\right) \delta \left( s-\left\vert E_{1}-E_{2}\right\vert \right) \text{,}
\end{equation}%
where $N$ is the number of levels in $\left\{ E_{i}\right\} $ and the
analytical result is quite complicated for general $N$. Instead, Wigner
proposes a surmise that we can focus on the $N=2$ case, the distribution
then reduces to{\small
\begin{equation}
P\left( s\right) \propto \int_{-\infty }^{\infty }\left\vert
E_{1}-E_{2}\right\vert ^{\nu }\delta \left( s-\left\vert
E_{1}-E_{2}\right\vert \right) e^{-A\sum_{i}E_{i}^{2}}dE_{1}dE_{2}\text{.}
\end{equation}%
} By introducing $x_{1}=E_{1}-E_{2}$, $x_{2}=E_{1}+E_{2}$, we have%
\begin{eqnarray}
P\left( s\right) &\propto &2\int_{-\infty }^{\infty }\left\vert
x_{1}\right\vert ^{\nu }\delta \left( s-\left\vert x_{1}\right\vert \right)
e^{-\frac{A}{2}\sum_{i}x_{i}^{2}}dx_{1}dx_{2}  \notag \\
&=&Cs^{\nu }e^{-As^{2}/2}\text{.}
\end{eqnarray}%
The constants $A,C$ can be determined by working out the integral about $%
x_{2}$, but it is more convenient to obtain by imposing the normalization
condition%
\begin{equation}
\int_{0}^{\infty }P\left( s\right) ds=1\text{, }\int_{0}^{\infty }sP\left(
s\right) ds=1\text{.}  \label{equ:normalization}
\end{equation}%
From which we can reach to the celebrated Wigner-Dyson distribution
\begin{equation}
P(s)=\left\{
\begin{array}{ll}
\frac{\pi }{2}s\exp \big(-\frac{\pi }{4}s^{2}\big) & \nu =1\quad \text{GOE}
\\[1mm]
\frac{32}{\pi ^{2}}s^{2}\exp \big(-\frac{4}{\pi }s^{2}\big) & \nu =2\quad
\text{GUE} \\[1mm]
\frac{2^{18}}{3^{6}\pi ^{3}}s^{4}\exp \big(-\frac{64}{9\pi }s^{2}\big) & \nu
=4\quad \text{GSE}%
\end{array}%
\right.  \label{equ:nearest}
\end{equation}

On the other hand, the levels are independent in Poisson ensemble, which
means the occurrence of next level is independent of previous level, the
nearest level spacings then follows a Poisson distribution $P\left( s\right)
=\exp \left( -s\right) $.

Although the Wigner surmise is for $2\times 2$ matrix, it works fairly good
when the matrix dimension is large. To demonstrate this, we present
numerical evidence from a quantum many-body system -- the spin-$1/2$
Heisenberg model with random external field, which is the canonical model in
the study of many-body localization (MBL), whose Hamiltonian in a length-$L$
chain is%
\begin{equation}
H=\sum_{i=1}^{L}\mathbf{S}_{i}\cdot \mathbf{S}_{i+1}+\sum_{i=1}^{L}\sum_{%
\alpha =x,y,z}h^{\alpha }\varepsilon _{i}^{\alpha }S_{i}^{\alpha },
\label{equ:H}
\end{equation}%
where we set coupling strength to be $1$ and assume periodic boundary
condition in Heisenberg term. The $\varepsilon _{i}^{\alpha }$'s are random
numbers within range $\left[ -1,1\right] $, and $h^{\alpha }$ is referred as
the randomness strength. We focus on two choices of $h^{\alpha }$: (i) $%
h^{x}=h^{z}=h\neq 0$ and $h^{y}=0$, the Hamiltonian matrix is orthogonal;
(ii) $h^{x}=h^{y}=h^{z}=h\neq 0$, the model being unitary. This model
undergos a thermal-MBL transition at roughly $h_{c}\simeq 3$ ($2.5$) in the
orthogonal (unitary) model, where the level spacing distribution evolves
from GOE (GUE) to Poisson\cite{Regnault16}.

\begin{figure}[htbp]
\includegraphics[width=\columnwidth]{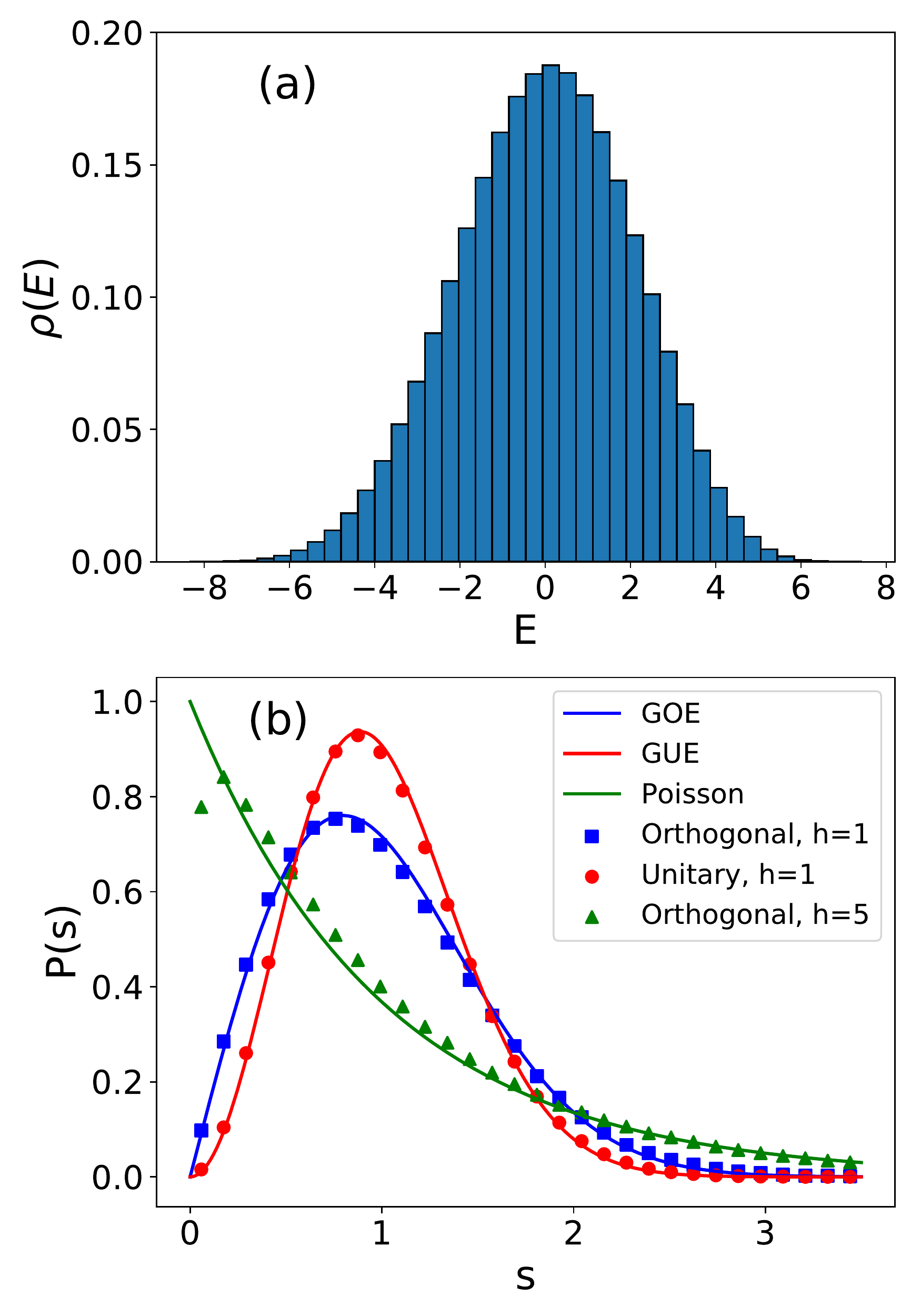}
\caption{(a) The density of states (DOS) $\protect\rho (E)$ of random field
Heisenberg model at $L=12$ and $h=1$ in orthogonal case, the DOS is more uniform in the middle
part, we therefore choose the middle half levels to do level statistics. (b)
Distribution of nearest level spacings $P(E_{i+1}-E_{i})$, we see a GOE/GUE
distribution for $h=1$ in the orthogonal/unitary model,
while a Poisson distribution is found for $h=5$ in orthogonal model,
the result for $h=5$ in unitary model is not displayed since it
coincides with that in the orthogonal model.} \label{fig:NN_spacing}
\end{figure}

We choose a $L=12$ system to present a numerical simulation, and prepare $%
500 $ samples at $h=1$ and $h=5$ for both the orthogonal and unitary model.
In Fig.~\ref{fig:NN_spacing}(a) we plot the density of states (DOS) for the $%
h=1 $ case in orthogonal model. We can see DOS is much more uniform in the
middle part of the spectrum, which is also the case for $h=5$ and unitary
model. Therefore we choose the middle half of energy levels to do the
spacing counting, and the results are shown in Fig.~\ref{fig:NN_spacing}(b).
We observe a clear GOE/GUE distribution for $h=1$ in orthogonal/unitary
model and a Poisson distribution for $h=5$ in orthogonal model as expected,
the fitting result for $h=5$ in unitary model is not shown since it almost
coincides with that in orthogonal model. It is noted the fitting for Poisson
distribution has minor deviations around the region $s\sim 0$, this is due
to finite size effect since there will always remain exponentially-decaying
but finite correlation between levels in a finite system. As we will
demonstrate in subsequent section, the fitting for higher order level
spacings will be better since the overlap between levels decays
exponentially with their distance in MBL phase.

A technique issue is, when counting the level spacings, we choose to take
the middle half levels of the spectrum, while we can also employ a unfolding
procedure using a spline interpolation that incorporates all energy levels%
\cite{Avishai2002}, and the fitting results are almost the same\cite%
{Regnault162,Rao182}.

\section{Higher Order Level Spacings}

Now we proceed to consider the distribution of higher order level spacings $%
\left\{ s_{i}^{\left( n\right) }=E_{i+n}-E_{i}\right\} $, using a
Wigner-like surmise. We first give the analytical derivation, then provide
numerical evidence from simulation of spin model in Eq.~(\ref{equ:H}) as
well as the non-trivial zeros of Riemann zeta function.

\subsection{Analytical Derivation}

\label{analytical}

Introduce $P_{n}\left( s\right) =P\left( s^{\left( n\right) }=s\right)
\equiv P\left( \left\vert E_{i+n}-E_{i}\right\vert =s\right) $, to apply the
Wigner surmise, we are now considering $\left( n+1\right) \times \left(
n+1\right) $ matrices, the distribution $P_n(s)$ then goes to%
\begin{eqnarray}
P_{n}\left( s\right) &\propto &\int_{-\infty }^{\infty
}\prod_{i<j}\left\vert E_{i}-E_{j}\right\vert ^{\nu }\delta \left(
s-\left\vert E_{1}-E_{n+1}\right\vert \right)  \notag \\
&&\times e^{-A\sum_{i=1}^{n+1}E_{i}^{2}}\prod_{i=1}^{n+1}dE_{i}
\end{eqnarray}%
We first change the variables to%
\begin{equation}
x_{i}=E_{i}-E_{i+1}\text{, }i=1,2,...,n\text{; }\quad
x_{n+1}=\sum_{i=1}^{n+1}E_{i}\text{,}
\end{equation}%
the $P_{n}\left( s\right) $ then evolves into%
\begin{widetext}
\begin{equation}
P_n\left( s \right) \propto \int_{-\infty }^{\infty
}\frac{\partial \left( E_{1},E_{2},...,E_{n+1}\right) }{\partial \left(
x_{1},x_{2},...,x_{n+1}\right) }\left( \prod_{i=1}^{n}\prod_{j=i}^{n}\left\vert
\sum_{k=i}^{j}x_{k}\right\vert ^{\nu }\right) \delta \left( s-\left\vert
\sum_{i=1}^{n}x_{i}\right\vert \right)
e^{-\frac{A}{n}\left[ \sum_{i=1}^{n}%
\sum_{j=i}^{n}\left( \sum_{k=i}^{j}x_{k}\right) ^{2}+x_{n+1}^{2}\right] }\prod_{i=1}^{n+1}dx_{i}.
\end{equation}%
\end{widetext}In this expression, the Jacobian $\frac{\partial \left(
E_{1},E_{2},...,E_{n+1}\right) }{\partial \left(
x_{1},x_{2},...,x_{n+1}\right) }$ and integral for $x_{n+1}$ are all
constants that can be absorbed into the normalization factor, hence we can
simplify $P_{n}\left( s\right) $ to%
\begin{eqnarray}
P_{n}\left( s\right) &\propto &\int_{-\infty }^{\infty }\left(
\prod_{i=1}^{n}\prod_{j=i}^{n}\left\vert \sum_{k=i}^{j}x_{k}\right\vert
^{\nu }\right) \delta \left( s-\left\vert \sum_{i=1}^{n}x_{i}\right\vert
\right)  \notag \\
&&\times e^{-\frac{A}{n}\sum_{i=1}^{n}\sum_{j=i}^{n}\left(
\sum_{k=i}^{j}x_{k}\right) ^{2}}\prod_{i=1}^{n}dx_{i}.
\end{eqnarray}%
Next, we introduce the $n$-dimensional spherical coordinate
\begin{eqnarray}
x_{1} &=&r\cos \theta _{1}\text{; }\quad x_{n}=r\prod_{i=1}^{n-1}\sin \theta
_{i}\text{;}  \notag \\
x_{i} &=&r\left( \prod_{j=1}^{i-1}\sin \theta _{j}\right) \cos \theta _{i}%
\text{, \thinspace }i=2,3,...,n-1\text{;} \\
0 &\leq &\theta _{i}\leq \pi \text{, }i=1,2,...,n-2\text{;}\quad 0\leq
\theta _{n-1}\leq 2\pi \text{,}  \notag
\end{eqnarray}%
whose Jacobian is%
\begin{equation}
\frac{\partial \left( x_{1},x_{2},...,x_{n}\right) }{\partial \left(
r,\theta _{1},\theta _{2},...,\theta _{n-1}\right) }=r^{n-1}%
\prod_{i=1}^{n-2}\sin ^{n-1-i}\theta _{i}  \label{equ:Jac}
\end{equation}%
which reduces to the normal spherical coordinate when $n=3$. The resulting
expression of $P_{n}\left( s\right) $ is complicated, while we are mostly
interested in the scaling behavior about $s$, hence we can write the formula
as%
\begin{eqnarray}
P_{n}\left( s\right) &\propto &\int_{0}^{\infty }r^{n-1}\int r^{\nu
C_{n+1}^{2}}\delta \left( s-r\left\vert G\left( \boldsymbol{\theta }\right)
\right\vert \right)  \notag \\
&&\times H\left( \boldsymbol{\theta }\right) e^{-\frac{A}{n}r^{2}J\left(
\boldsymbol{\theta }\right) }drd\boldsymbol{\theta }
\end{eqnarray}%
where $C_{n+1}^{2}=n\left( n+1\right) /2$, and $d\boldsymbol{\theta }$ $%
=\prod_{i=1}^{n-1}d\theta _{i}$, the explanation goes as follows: (i) the
first term $r^{n-1}$ comes from the radial part of the Jacobian in Eq.~(\ref%
{equ:Jac}); (ii) the second $r^{\nu C_{n+1}^{2}}$ comes number of terms in $%
\prod_{i=1}^{n}\prod_{j=i}^{n}\left\vert \sum_{k=i}^{j}x_{i}\right\vert
^{\nu }$, where each term contributes a factor $r^{\nu }$; (iii) the
auxiliary function $G\left( \boldsymbol{\theta }\right)
=\sum_{i=1}^{n}x_{i}/r$; (iv) the second auxiliary function $H\left(
\boldsymbol{\theta }\right) $ is comprised of the angular part of the
Jacobian and the angular part of $\prod_{i=1}^{n}\prod_{j=i}^{n}\left\vert
\sum_{k=i}^{j}x_{i}\right\vert ^{\nu }$; (v) $J\left( \boldsymbol{\theta }%
\right) $ is the angular part of $\sum_{i=1}^{n}\sum_{j=i}^{n}\left(
\sum_{k=i}^{j}x_{k}\right) ^{2}$. The key observation is that $G\left(
\boldsymbol{\theta }\right) ,H\left( \boldsymbol{\theta }\right) ,J\left(
\boldsymbol{\theta }\right) $ all depend only on $\boldsymbol{\theta }$
while independent of $r$. Since we are only interested in the scaling
behavior about $s$, we can work out the delta function, and get%
\begin{equation}
P_{n}\left( s\right) \propto s^{\nu C_{n+1}^{2}+n-1}\int H\left( \boldsymbol{%
\theta }\right) e^{-\frac{AJ\left( \boldsymbol{\theta }\right) }{n\left\vert
G\left( \boldsymbol{\theta }\right) \right\vert ^{2}}s^{2}}d\boldsymbol{%
\theta }
\end{equation}%
Although the integral for $\boldsymbol{\theta }$ is tedious and difficult to
handle, it will only make correction to the Gaussian factor while not
influence the scaling behavior about $s$. Therefore we can write $%
P_{n}\left( s\right) $ into a generalized Wigner-Dyson distribution%
\begin{eqnarray}
P_{n}\left( s\right) &=&C\left( \alpha \right) s^{\alpha }e^{-A\left( \alpha
\right) s^{2}}\text{, }  \label{equ:GWD} \\
\alpha &=&\frac{n\left( n+1\right) }{2}\nu +n-1\text{.}  \label{equ:rescale}
\end{eqnarray}%
The normalization factors $C\left( \alpha \right) $ and $A\left( \alpha
\right) $ can be determined by the normalization condition in Eq.~(\ref%
{equ:normalization}), for which we obtain%
\begin{equation}
A\left( \alpha \right) =\left( \frac{\Gamma \left( \alpha /2+1\right) }{%
\Gamma \left( \alpha /2+1/2\right) }\right) ^{2}\text{, }C\left( \alpha
\right) =\frac{2\Gamma ^{\alpha +1}\left( \alpha /2+1\right) }{\Gamma
^{\alpha +2}\left( \alpha /2+1/2\right) }\text{,}
\end{equation}%
where $\Gamma \left( z\right) =\int_{0}^{\infty }t^{z-1}e^{-t}dt$ is the
Gamma function. When $n=1$, $P_n(s) $ reduces to the conventional
Wigner-Dyson distribution in Eq.~(\ref{equ:nearest}).

Interestingly, there exists coincidence between distributions in different
ensembles. For example, as has been known for a long time\cite{Mehta,GSE}, $%
P_{k}\left( s\right) $ in the GSE coincides with $P_{2k}\left( s\right) $ in
GOE for arbitrary integer $k$. And $P_{7}\left( s\right) $ in GOE coincides
with $P_{5}\left( s\right) $ in GUE, and so on. Actually, our derivations
are purely mathematical that works for arbitrary positive values of $\nu $
(not limited to integer values), although the three standard Gaussian
ensembles are of most physical interest.

For the uncorrelated energy levels in the Poisson class, the distribution
for higher order spacing can also be obtained. Let's start with $n=2$, we
can write $\widetilde{s}=E_{i+2}-E_{i}=\left( E_{i+2}-E_{i+1}\right) +\left(
E_{i+1}-E_{i}\right) =s_{i+1}+s_{i}$, where $s_{i+1}$ and $s_{i}$ can be
treated as independent variables that both follows Poisson distribution,
therefore the distribution $P_{2}\left( \widetilde{s}\right) $ for
unnormalized $\widetilde{s}$\ is%
\begin{equation}
P\left( \widetilde{s}\right) \propto \int_{0}^{\widetilde{s}}P_{1}\left(
\widetilde{s}-s_{1}\right) P_{1}\left( s_{1}\right) ds_{1}=\widetilde{s}e^{-%
\widetilde{s}}\text{.}  \label{equ:recur}
\end{equation}%
Then by requiring the normalization condition\ we arrive at $P_{2}\left(
s\right) =4se^{-2s}$ -- the semi-Poisson distribution\cite{semiPoisson},
which is suggested to be the distribution for nearest level spacing at the
thermal-MBL transition point in orthogonal model \cite{Serbyn}. This
interesting fact indicates the (leading order) universality of this
transition point is more affected by the MBL phase rather than the thermal
phase, which is already noticed by previous studies\cite{Huse2,Serbyn}.

For higher order level spacing in Poisson ensemble, by repeating the
procedure in Eq.~(\ref{equ:recur}) $n-1\ $times, we reach to%
\begin{equation}
P_{n}\left( s\right) =\frac{n^{n}}{\left( n-1\right) !}s^{n-1}e^{-ns}\text{.}
\label{equ:Pn}
\end{equation}%
which is a generalized semi-Poisson distribution with index $n$. Compared to
the Poisson distribution for nearest level spacings, it's crucial to note
that $P_{n}\left( 0\right) =0$ for $n\geq 2$, this is not a result of level
repulsion as in the Gaussian ensembles, rather, it simply states that $%
n+1\left( n\geq 2\right) $ consecutive levels do not coincide.

We note every $P_{n}\left( s\right) $ in the Gaussian and Poisson ensembles
tends to be the Dirac delta function $\delta \left( s-1\right) $ in the
limit $n\rightarrow \infty $, which is easily understood since in that limit
only one spacing remains in the spectrum. Finally, we want to emphasize that
the levels are well-correlated in the Gaussian ensembles, hence the
derivation of $P_{n}\left( s\right) $ for Poisson ensemble in Eq.~(\ref%
{equ:recur}) do not hold, otherwise the result will deviate dramatically\cite%
{Rubah}.

For convenience we list the order of the polynomial part in $P_{n}\left(
s\right) $ for the three Gaussian ensembles as well as Poisson ensemble up
to $n=8$ in Table~\ref{tab:1}, note that the exponential parts in the former
class are Gaussian type and that for Poisson ensemble is a exponential decay.

\begin{table}
\centering

\begin{tabular}{|c|c|c|c|c|c|c|c|c|}
\hline
$n$ & $1$ & $2$ & $3$ & $4$ & $5$ & $6$ & $7$ & $8$ \\ \hline
GOE & $1$ & $4$ & $8$ & $13$ & $19$ & $26$ & $34$ & $43$ \\ \hline
GUE & $2$ & $7$ & $14$ & $23$ & $34$ & $47$ & $62$ & $79$ \\ \hline
GSE & $4$ & $13$ & $26$ & $43$ & $64$ & $89$ & $118$ & $151$ \\ \hline\hline
Poisson & $0$ & $1$ & $2$ & $3$ & $4$ & $5$ & $6$ & $7$ \\ \hline
\end{tabular}%
\caption{The order of the polynomial term in
$P_{n}(s)$ for the three Gaussian ensembles as well as Poisson ensemble,
the decaying term is Gaussian type for the former class and exponential decay
for the latter.} \label{tab:1}
\end{table}

\subsection{Numerical Simulation}

\label{numerics}

\begin{figure}[htbp]
\includegraphics[width=\columnwidth]{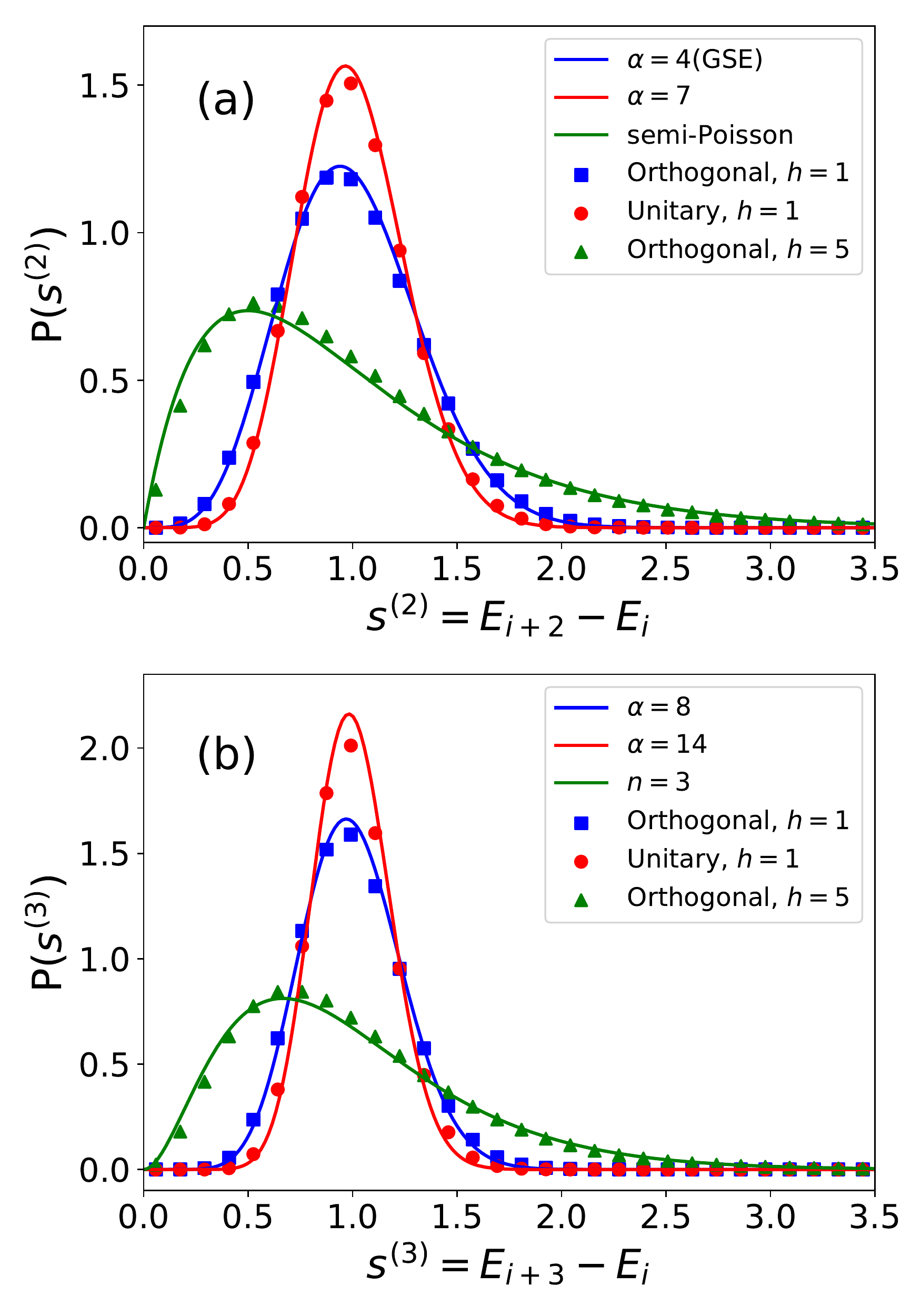}
\caption{Distribution of next-nearest level spacings $P(s^{(2)})$ in (a)
and next-next-nearest level spacings $P(s^{(3)})$ in (b), where
$\alpha$ and $n$ are the index in Eq.~(\ref{equ:GWD}) and Eq.~(\ref{equ:Pn}) respectively.}
\label{fig:higher_spacing}
\end{figure}

To show how well the distributions in Eq.~(\ref{equ:GWD}) and Eq.~(\ref%
{equ:Pn}) work for matrix with large dimension, we now perform numerical
simulations for the random spin model in Eq.~(\ref{equ:H}), where we also
pick the middle half levels to do statistics. We have tested the formula up
to $n=5$, and in Fig.~\ref{fig:higher_spacing} we display the fitting
results for $n=2$ and $n=3$.

As expected, the fittings are quite accurate for both GOE and GUE as well as
Poisson ensemble. In fact, the fittings for higher order spacings in the
Poisson ensemble are better than that for nearest spacing in Fig.~\ref%
{fig:NN_spacing}(b). This is because in MBL phase the overlap between levels
decays exponentially with their distance, hence the fitting for higher order
level spacings is less affected by finite size effect.

For another example we consider the non-trivial zeros of the Riemann zeta
function\cite{zeta}%
\begin{equation}
\zeta \left( z\right) =\sum_{n=1}^{\infty }\frac{1}{n^{z}}\text{,}
\end{equation}%
it was established that statistical properties of non-trivial Riemann zeros $%
\left\{ \gamma _{i}\right\} $ are well described by the GUE distribution\cite%
{Zeta}. Therefore, we expect the gaps $\left\{ s_{i}^{(n)}=\gamma
_{i+n}-\gamma _{i}\right\} $ follows the same distribution as those in GUE.
The numerical results for $n=1,2,3$ are presented in Fig.~\ref{fig:zeta}, as
can be seen, the fittings are perfect.

\begin{figure}[htbp]
\includegraphics[width=\columnwidth]{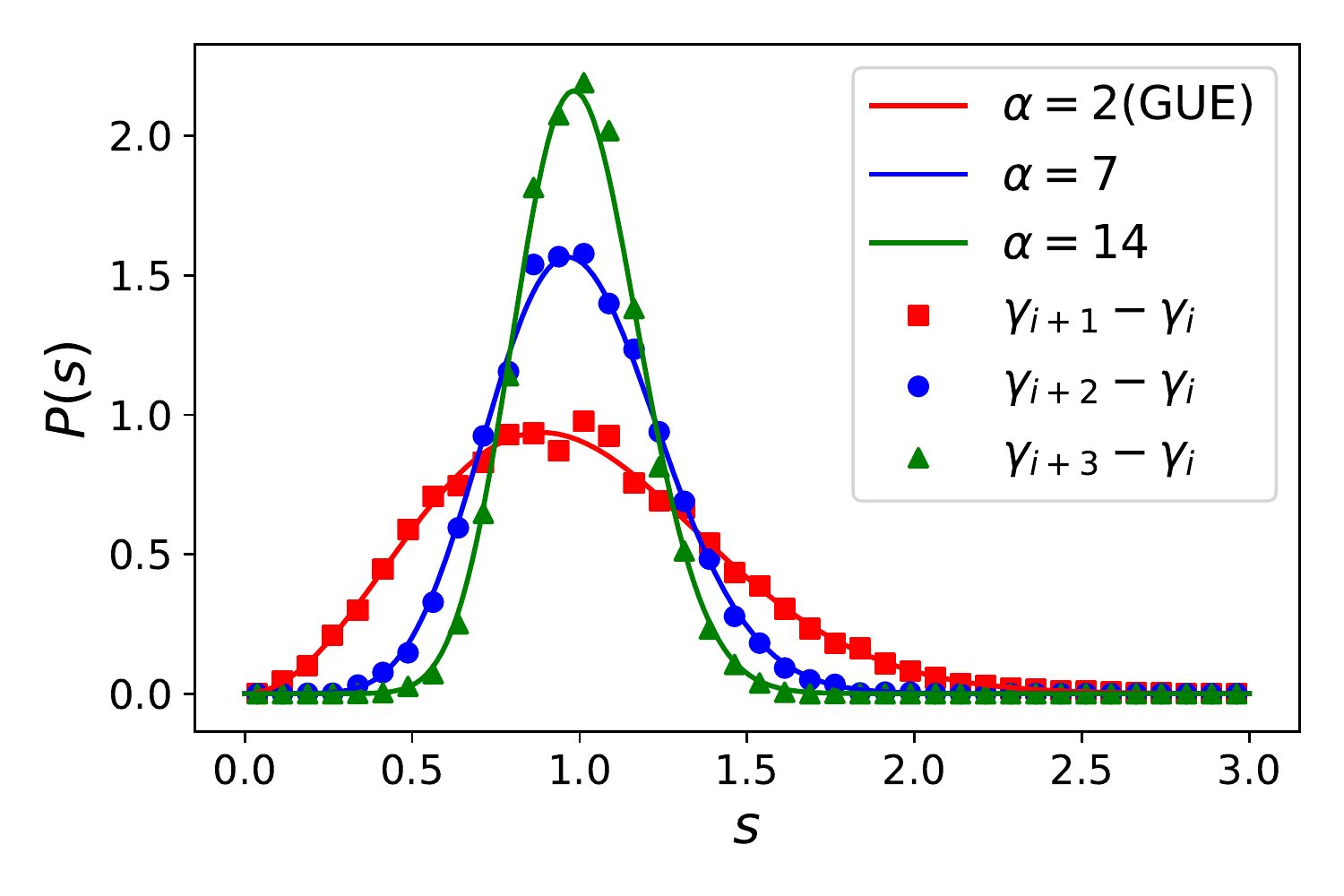}
\caption{The distribution of $n$-th order spacings of the non-trivial
zeros $\{\gamma_i\}$ of Riemann zeta function, where $\alpha$ is the index in
generalized Wigner-Dyson distribution in Eq.~(\ref{equ:GWD}).
The data comes from $10^4$ levels starting
from the $10^{22}$th zero, taken from Ref.~[\onlinecite{Odlyzko}].} \label%
{fig:zeta}
\end{figure}

\section{Higher Order Gap Ratios}

\label{ratio} As mentioned in Sec.~\ref{intro}, besides the level spacings,
another quantity is also widely used in the study of random matrices, namely
the ratio between adjacent gaps $\left\{ r_{i}=s_{i+1}/s_{i}\right\} $,
which is independent of local DOS. The distribution of nearest gap ratios $%
P\left( \nu ,r\right) $ is given in Ref.~[\onlinecite{Atas}], whose result is%
\begin{equation}
P\left( \nu ,r\right) =\frac{1}{Z_{\nu }}\frac{\left( r+r^{2}\right) ^{\nu }%
}{\left( 1+r+r^{2}\right) ^{1+3\nu /2}}
\end{equation}%
where $\nu =1,2,4$ for GOE,GUE,GSE, and $Z_{\nu }$ is the normalization
factor determined by requiring $\int_{0}^{\infty }P\left( \nu ,r\right) dr=1$%
.

\begin{figure}[htbp]
\includegraphics[width=8cm]{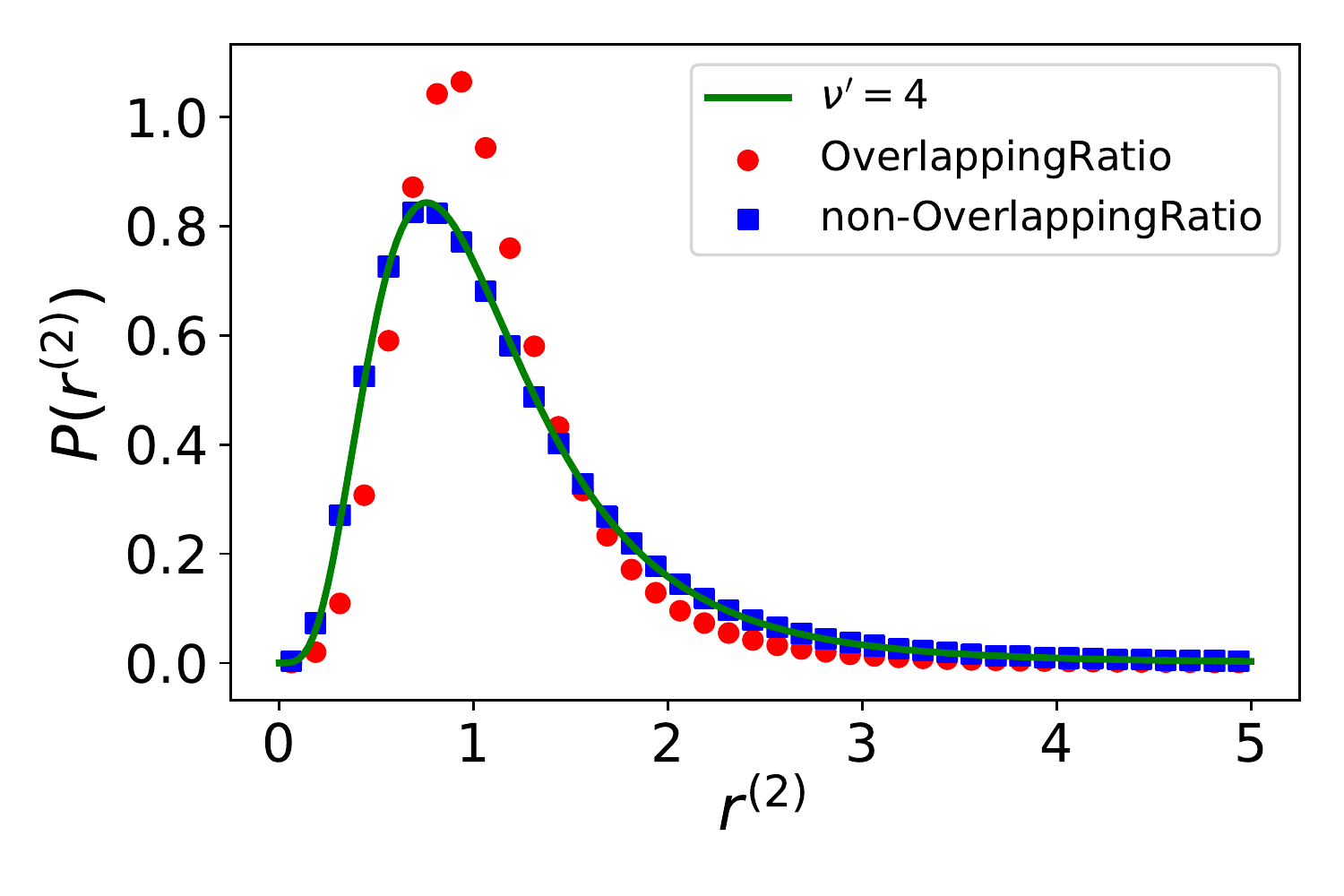}
\caption{The distribution of second-order gap ratio in the orthogonal model,
where red and blue dots correspond to overlapping and non-overlapping ratios respectively,
the latter fits perfectly with the formula in Eq.~(\ref{equ:rn}) with $\nu^{\prime}=4$.
Note the data is taken from the whole energy spectrum without unfolding.} %
\label{fig:ratiocom}
\end{figure}

This gap ratio can also be generalized to higher order, but in different
ways, i.e. the \textquotedblleft overlapping\textquotedblright\ \cite%
{Atas,Atas2} and \textquotedblleft non-overlapping\textquotedblright\ \cite%
{Tekur,Chavda} way. In the former case we are dealing with%
\begin{equation}
\widetilde{r}_{i}^{\left( n\right) }=\frac{E_{i+n}-E_{i}}{E_{i+n-1}-E_{i-1}}=%
\frac{s_{i+n}+s_{i+n-1}+...+s_{i+1}}{s_{i+n-1}+s_{i+n-2}+...+s_{i}}\text{,}
\end{equation}%
which is named \textquotedblleft overlaping\textquotedblright\ ratio since
there is shared spacings between the numerator and denominator. While the
\textquotedblleft non-overlapping\textquotedblright\ ratio is defined as%
\begin{equation}
r_{i}^{\left( n\right) }=\frac{E_{i+2n}-E_{i+n}}{E_{i+n}-E_{i}}=\frac{%
s_{i+2n}+s_{i+2n-1}+...+s_{i+n+1}}{s_{i+n}+s_{i+n-1}+...+s_{i}}\text{.}
\end{equation}%
Both these two generalizations reduce to the nearest gap ratio when $n=1$,
but they are quite different when studying their distributions using Wigner
surmise: for overlapping ratio $\widetilde{r}_{i}^{\left( n\right) }$, the
smallest matrix dimension is $\left( n+2\right) \times \left( n+2\right) $;
while it is $\left( 1+2n\right) \times \left( 1+2n\right) $ for
non-overlapping ratio. Naively, we can
expect the distribution for $\widetilde{r}^{\left( n\right) }$ is more
involved due to the overlapping spacings. Indeed, the $n=2$ case for $%
P\left( \widetilde{r}^{\left( n\right) }\right) $ has been worked out in
Ref.~[\onlinecite{Atas2}] and the result is very complicated. Instead, for the
non-overlapping ratio, Ref.~[\onlinecite{Tekur}] provides
compelling numerical evidence for its distribution to follow
\begin{eqnarray}
P\left( \nu ,r^{\left( n\right) }\right)  &=&P\left( \nu ^{\prime },r\right)
\text{, }  \label{equ:rn} \\
\nu ^{\prime } &=&\frac{n\left( n+1\right) }{2}\nu +n-1\text{.}
\label{equ:rescale2}
\end{eqnarray}%
Surprisingly, the rescaling relation Eq.~(\ref{equ:rescale2}) coincides with
that for higher order level spacing in Eq.~(\ref{equ:rescale}). We have also
confirmed this formula by numerical simulations in our spin model Eq.~(\ref%
{equ:H}), and the results for $n=2$ in GOE ($\nu =1$) case is presented in
Fig.~\ref{fig:ratiocom}, where we also draw the distribution of overlapping
ratio $\widetilde{r}^{\left( 2\right) }$ for comparison. As can be seen,
they differ dramatically, and the fitting for non-overlapping ratio is quite
accurate. This result strongly suggest the non-overlapping ratio is more
universal than the overlapping ratio, and its distribution $P\left(
r^{\left( n\right) }\right) $ is homogeneously related with that for the $n-$th
order level spacing, at least in the sense of Wigner surmise, for which
we provide a heuristic explanation as follows.

For a given energy spectrum $\left\{ E_{i}\right\} $ from a Gaussian
ensemble with index $\nu $, we can make up a new spectrum $\left\{
E_{i}^{^{\prime }}\right\} $ by picking one level from every $n$ levels in $%
\left\{ E_{i}\right\} $, then the $n$-th order level spacing $s^{\left(
n\right) }$ in $\left\{ E_{i}\right\} $ becomes the nearest level spacing in
$\left\{ E_{i}^{^{\prime }}\right\} $, and the $n$-th order non-overlapping
ratio in $\left\{ E_{i}\right\} $ becomes the nearest gap ratio in $\left\{
E_{i}^{^{\prime }}\right\} $. Since we have analytically proven the
rescaling relation in Eq.~(\ref{equ:rescale}), we conjecture the probability
density for $\left\{ E_{i}^{^{\prime }}\right\} $ (to leading order) bear
the same form as $\left\{ E_{i}\right\} $ in Eq.~(\ref{equ:Dist}) with the
rescaled parameter $\alpha $ in Eq.~(\ref{equ:rescale}). Therefore, the
higher order non-overlapping gap ratios also follow the same rescaling as
expressed in Eq.~(\ref{equ:rn}) and Eq.~(\ref{equ:rescale2}). For this point
of view, numerical evidences are provided in a recent work of the author\cite%
{Rao20}.

\section{Conclusion and Discussion}

\label{conclusion}

We have analytically studied the distribution of higher order level spacings
$\left\{ s_{i}^{\left( n\right) }=E_{i+n}-E_{i}\right\} $ which describes
the level correlations on long range. It is shown $s^{\left( n\right) }$ in
the Gaussian ensemble with index $\nu $ follows a generalized Wigner-Dyson
distribution with index $\alpha =\nu C_{n+1}^{2}+n-1$, where $\nu =1,2,4$
for GOE,GUE,GSE respectively. This results in a large number of coincident
relations for distributions of level spacings of different orders in
different ensembles. While $s^{\left( n\right) }$ in Poisson ensemble follows
a generalized semi-Poisson distribution with index $n$. Our derivation is
rigorous based on a Wigner-like surmise, and the results have been confirmed
by numerical simulations from random spin system and non-trivial zeros of
Riemann zeta function.

We also discussed the higher order generalization of gap ratios, which come
in two different ways -- the ``overlapping'' and ``
non-overlapping'' way -- and point out their difference in
studying their distributions using Wigner-like surmise. Notably, the
distribution for the non-overlapping gap ratio has been studied numerically
in Ref.~[\onlinecite{Tekur}], in which the authors find a scaling relation
Eq.~(\ref{equ:rescale2}) that is identical to the one we find analytically
for higher order level spacings. This strongly indicates the distribution of
higher order spacing and non-overlapping gap ratio is correlated in a
homogeneous way, for which we provided a heuristic explanation.

It's noted the higher-order level spacings have played an important role in
the study of the spacing distribution in a spectrum with missing levels\cite{Bohigas},
where the second order level spacing distribution in GOE is derived by a
method different from this work. Our derivations for $P\left( s^{\left( n\right)
}\right) $ in Guassian ensembles are purely mathematical that work for
arbitrary positive values of $\nu $, although the $\nu =1,2,4$ for
GOE,GUE,GSE are of most physical interest. Therefore, it is possible for our
results to find applications in models that goes beyond the three standard Gaussian
ensembles. For example, the $\nu =3$ behavior for level spacing has been
found in a 2D lattice with non-Hermitian disorder\cite{Tzortzakakis}.

It is also interesting to note the distribution of next-nearest level
spacing in Poisson class is semi-Poisson $P_{2}\left( s\right) \propto s\exp
\left( -2s\right) $, which is suggested to be the distribution for nearest
level spacing at the thermal-MBL transition point in orthogonal model \cite%
{Serbyn}. This indicates -- to leading order -- the universality property of
this transition point is more affected by the MBL phase than the thermal
phase, a fact already noticed by previous studies\cite{Huse2,Serbyn}.
This observation thus motivates a natural question: how will the
thermal phase affect the universality of the MBL transition point?
To answer this question, a comparison between the GOE-Poisson and
GUE-Poisson transition points is suggested, which is left for
a future study.

Last but not least, in this paper
the distribution of higher order level spacing is derived only in $\left(
n+1\right) \times \left( n+1\right) $ matrix, its exact value in large
matrix as well as the difference between them can in principle be estimated
using the method in Ref.~[\onlinecite{Atas}], this is also left for a
future study.

\section*{Acknowledgements}

The author acknowledges the helpful discussions with Xin Wan and Rubah
Kausar. This work is supported by the National Natural Science Foundation of
China through Grant No.11904069 and No.11847005.

\end{document}